\documentclass[
               ]{jacow}

\usepackage[english]{babel}

\begin{document}

\title{Applicability Limits and Effectiveness of Canonical Angular Momentum as a Design Parameter in Arbitrary Magnetic Fields }

\author[,1]{H. Iinuma\thanks{hiromi.iinuma.spin@vc.ibaraki.ac.jp}}
\author[2,3]{R. Matsushita}
\author[3]{K. Sasaki}
\affil[1]{Ibaraki University, Mito, Japan}
\affil[2]{University of Tokyo, Tokyo, Japan}
\affil[3]{KEK, Tsukuba, Japan}

\maketitle

\begin{abstract}
Canonical quantities provide a natural framework for describing beam transport in spatially varying magnetic fields. In this study, the vector potential associated with a three-dimensional magnetic-field map is reconstructed using equivalent current elements determined by truncated singular-value decomposition. The reconstruction is adapted to the local magnetic-field symmetry and reproduces the reference field distribution along the representative injection trajectory. In the approximately axisymmetric region, the reconstructed vector potential is used to evaluate the canonical angular momentum. The inclusion of the vector-potential contribution substantially suppresses the variation of the mechanical angular momentum, resulting in an approximately constant canonical angular momentum. Extension to the non-axisymmetric injection region requires further examination of the consistency between independently reconstructed field regions.

\end{abstract}

\section{MOTIVATION}
The three-dimensional spiral injection has been investigated using a four-dimensional transverse phase-space representation in the beam coordinate system, where the \(Z\) axis is taken along the beam propagation direction and the transverse motion is represented by \((X,X',Y,Y')\) ~\cite{Iinuma:2016zfu}. 
This representation has also been employed in subsequent studies of the three-dimensional injection~\cite{Iinuma:ipac2025-wepm029,Ogawa:ipac2025-wepm055}.
However, the four-dimensional representation is based solely on mechanical momentum and does not incorporate the vector potential, and therefore does not explicitly represent canonical conserved quantities associated with the magnetic-field symmetry.
This observation motivated us to reconsider the formulation of the injection dynamics from the underlying Hamiltonian mechanics rather than assuming a reduced transverse phase space.

In a spatially varying magnetic field, the canonical momentum differs from the mechanical momentum and explicitly contains the vector potential \(\mathbf{A}\). Therefore, a canonical description of the three-dimensional particle motion requires the vector potential associated with the magnetic-field distribution. Based on this consideration, we have reformulated the three-dimensional spiral injection in the full canonical phase space~\cite{Iinuma:ipac2026-thp5620,Iinuma2026Canonical}. The present study focuses specifically on the reconstruction of the vector potential from the three-dimensional magnetic-field map and its application to the evaluation of canonical quantities along the injection trajectory.

An additional advantage of this formulation is that the particle dynamics can be treated directly in the laboratory coordinate system, without introducing a local beam coordinate system along an arbitrary three-dimensional reference trajectory. This eliminates a potential source of numerical uncertainty associated with defining and transforming the local coordinate frame. Although the canonical phase space is six-dimensional, retaining the intrinsic degrees of freedom provides a simpler and more direct description of the underlying dynamics.

\begin{figure}[!htb]
   \centering
      \includegraphics*[width=.7\columnwidth]{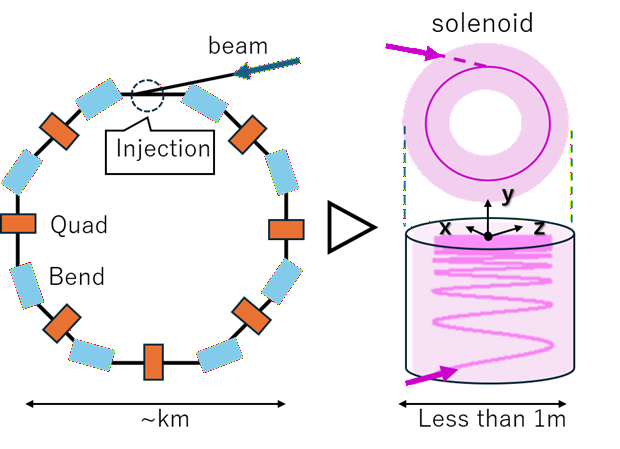}
   \caption{Schematic comparison between a conventional storage ring with a kilometer-scale circumference (left) and the compact storage ring considered in this study (right). In the compact ring, the injected beam undergoes three-dimensional spiral motion within the solenoidal magnetic field.}
   \label{fig:Image}
\end{figure}

The present study considers beam injection into a compact storage ring, as schematically illustrated in Fig.~\ref{fig:Image}. In conventional storage rings with circumferences typically on the kilometer scale, beam transport is generally designed as a sequence of localized magnetic elements separated by field-free regions. In contrast, the three-dimensional spiral injection considered here takes place within a compact solenoidal magnet with a characteristic size of less than one meter. 

The experimentally demonstrated three-dimensional spiral injection trajectory~\cite{PRL2026} provides a practical reference for investigating particle dynamics in the strongly varying fringe field of the storage magnet.
The injected beam therefore undergoes three-dimensional motion continuously in a spatially varying magnetic field. This requires a description of beam transport in the magnetic field itself, rather than one based primarily on field-free transport between localized magnetic elements.

\section{CANONICAL DESCRIPTION IN A MAGNETIC FIELD}
The three-dimensional spiral injection takes place in a spatially varying magnetic field, rather than in a field-free transport region. The dynamics of a charged particle in an external electromagnetic field is described by a Hamiltonian formulation in which the canonical momentum contains the vector potential \(\mathbf{A}\) [see, e.g., Jackson, Chap. 12, Sec. 12.1]~\cite{Jackson}. The canonical momentum is written as
\begin{equation}
\mathbf{P} = \mathbf{p}_{\mathrm{mech}} + q\mathbf{A}, 
\end{equation}

where \(\mathbf{p}_{\mathrm{mech}}\) is the mechanical momentum and \(q\) is the particle charge. Therefore, the vector potential associated with the magnetic-field distribution is required to construct the canonical phase space for the injection dynamics.

In the present study, the magnetic field is available as a three-dimensional field map, whereas the vector potential is not directly provided. The problem is therefore to reconstruct a vector potential \(\mathbf{A}\) satisfying
\begin{equation}
 \mathbf{B}=\nabla\times\mathbf{A}, 
\end{equation}

with sufficient accuracy for evaluating canonical quantities along the three-dimensional injection trajectory.

\section{VECTOR-POTENTIAL RECONSTRUCTION FROM THE 3D MAGNETIC-FIELD MAP}

\begin{figure}[!htb]
   \centering
   \includegraphics*[width=.32\columnwidth]{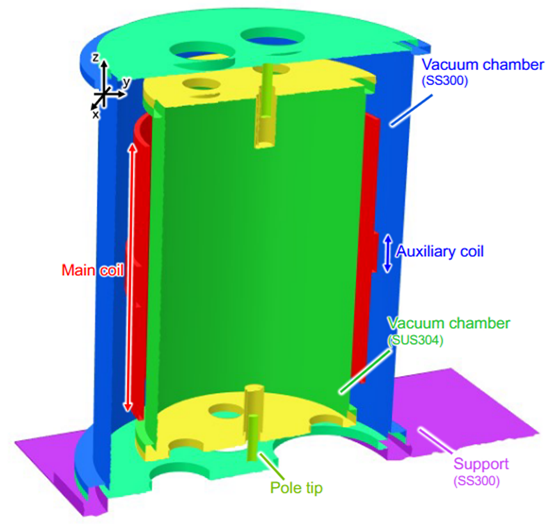}\qquad%
    \includegraphics*[width=.2\columnwidth]{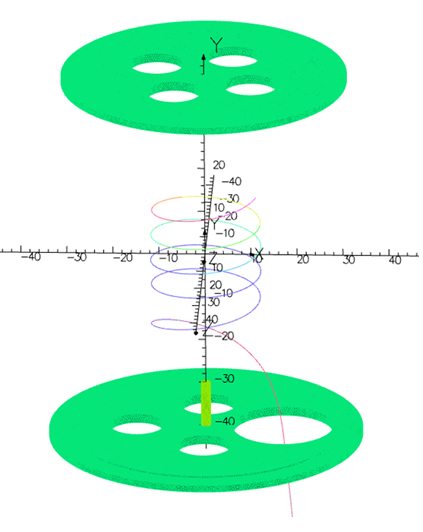}\qquad%
    \includegraphics*[width=.2\columnwidth]{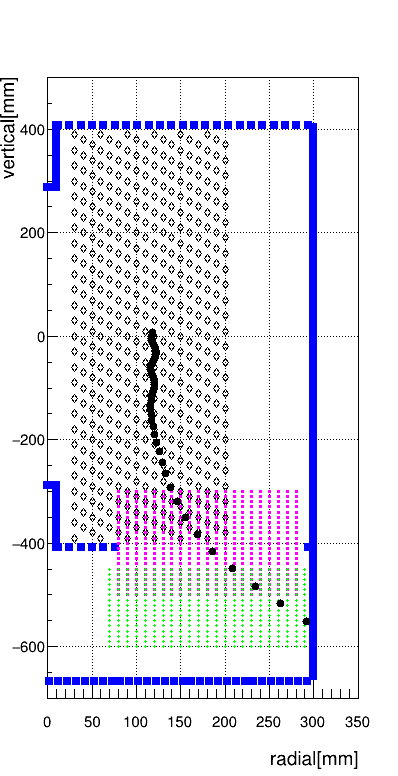}

   \caption{Magnetic-field model used for the vector-potential reconstruction. From left to right: the three-dimensional magnet model, the equivalent-current representation, and the reconstruction domain. The current amplitudes of the equivalent circular loops are determined to reproduce the reference magnetic-field map.}
   \label{fig:Model}
\end{figure}

The magnetic-field distribution of the injection region is obtained from a three-dimensional finite-element magnetic-field calculation~\cite{opera}. To reconstruct the vector potential from this field map, the magnet is represented by a set of equivalent circular current loops, as illustrated in Fig.~\ref{fig:Model}. The current amplitudes are determined so that the magnetic field generated by the equivalent-current model reproduces the reference magnetic-field map in the reconstruction domain. Since both the magnetic field and the vector potential are calculated from the same current distribution, the reconstructed vector potential is intrinsically consistent with the reconstructed magnetic field through \(\mathbf{B}=\nabla\times\mathbf{A}\).

\begin{figure}[!htb]
   \centering
   \includegraphics*[width=.4\columnwidth]{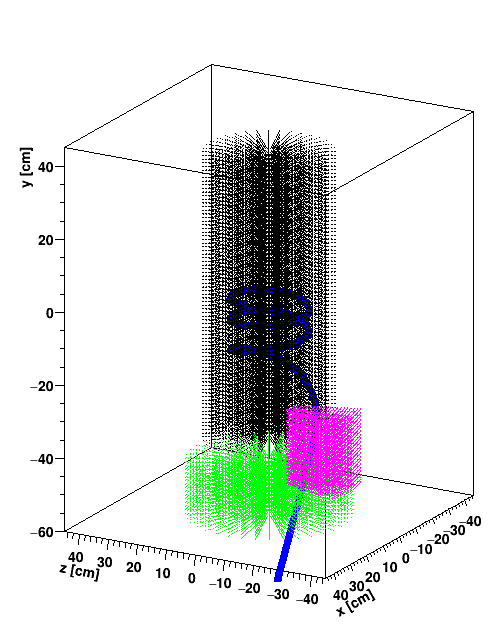}\qquad%
   \includegraphics*[width=.5\columnwidth]{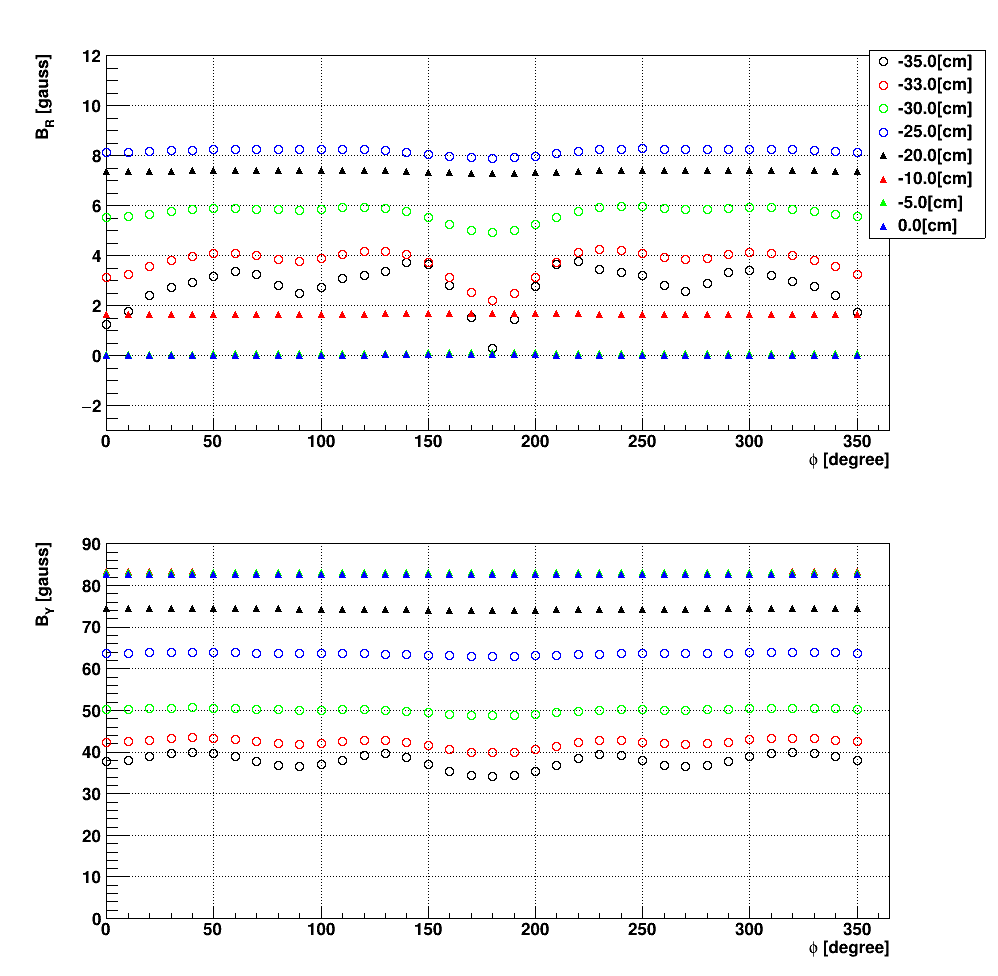}
  
   \caption{Reference magnetic-field map used for the reconstruction. Left: spatial distribution of the reference points and a representative injection trajectory. The black cylindrical region corresponds to the interior of the storage magnet, the magenta sector to the injection region, and the green cylindrical region to the region outside the iron yoke. Right: magnetic-field distribution at selected reference points inside the storage magnet as a function of the azimuthal angle. Different colors represent different positions along the solenoid axis \(Y\). The field is approximately axially symmetric near \(Y\simeq0\), whereas a systematic azimuthal dependence develops for \(Y<-0.3~\mathrm{m}\).}
   \label{fig:InnerField}
\end{figure}

The reference points of the three-dimensional magnetic-field map and a representative injection trajectory are illustrated in Fig.~\ref{fig:InnerField} (left). The field map is divided into three regions according to the magnetic configuration: a cylindrical region inside the storage magnet, shown in black; a sector-shaped injection region, shown in magenta; and a cylindrical region outside the iron yoke, shown in green.

Figure~\ref{fig:InnerField} (right) shows the magnetic field at selected reference points inside the storage magnet as a function of the azimuthal angle. Different colors correspond to different positions along the solenoid axis, \(Y\). Around the central region (\(Y\simeq0\)), the field distribution is nearly independent of the azimuthal angle, reflecting the axial symmetry of the solenoidal field. As the injection region is approached, particularly for \(Y<-0.3~\mathrm{m}\), a systematic azimuthal dependence appears, indicating the breakdown of axial symmetry. This spatial variation must therefore be taken into account when reconstructing the magnetic field and the associated vector potential in the injection region.

The equivalent currents are determined so that the magnetic field generated by the current elements reproduces the reference magnetic-field map shown in Fig.~\ref{fig:InnerField}. The current elements are placed in the region corresponding to the iron yoke, as indicated by the blue points in Fig.~\ref{fig:Model}. The relation between the magnetic field at the reference points and the equivalent currents is expressed as
\begin{equation}
 \mathbf{B}=\mathbf{M}\mathbf{I}, 
\end{equation}
    where \(\mathbf{B}\) represents the reference magnetic-field values, \(\mathbf{I}\) the equivalent-current amplitudes, and \(\mathbf{M}\) the response matrix calculated from the magnetic field generated by each current element. The equivalent currents are obtained by solving this inverse problem using truncated singular-value decomposition (tSVD)~\cite{Abe:NIMA890,oyama:MT}.

For the approximately axisymmetric region inside the iron yoke, the three-dimensional field map is first averaged over the azimuthal panels sampled at \(10^\circ\) intervals. The resulting field distribution is represented on a single \(R\)-\(Y\) plane using \(18 \times 79\) reference points in the radial and axial directions, respectively. The tSVD solution then determines the equivalent-current amplitudes at the current-element positions shown in Fig.~\ref{fig:Model}.

\begin{figure}[!htb]
   \centering
   \includegraphics*[width=.8\columnwidth]{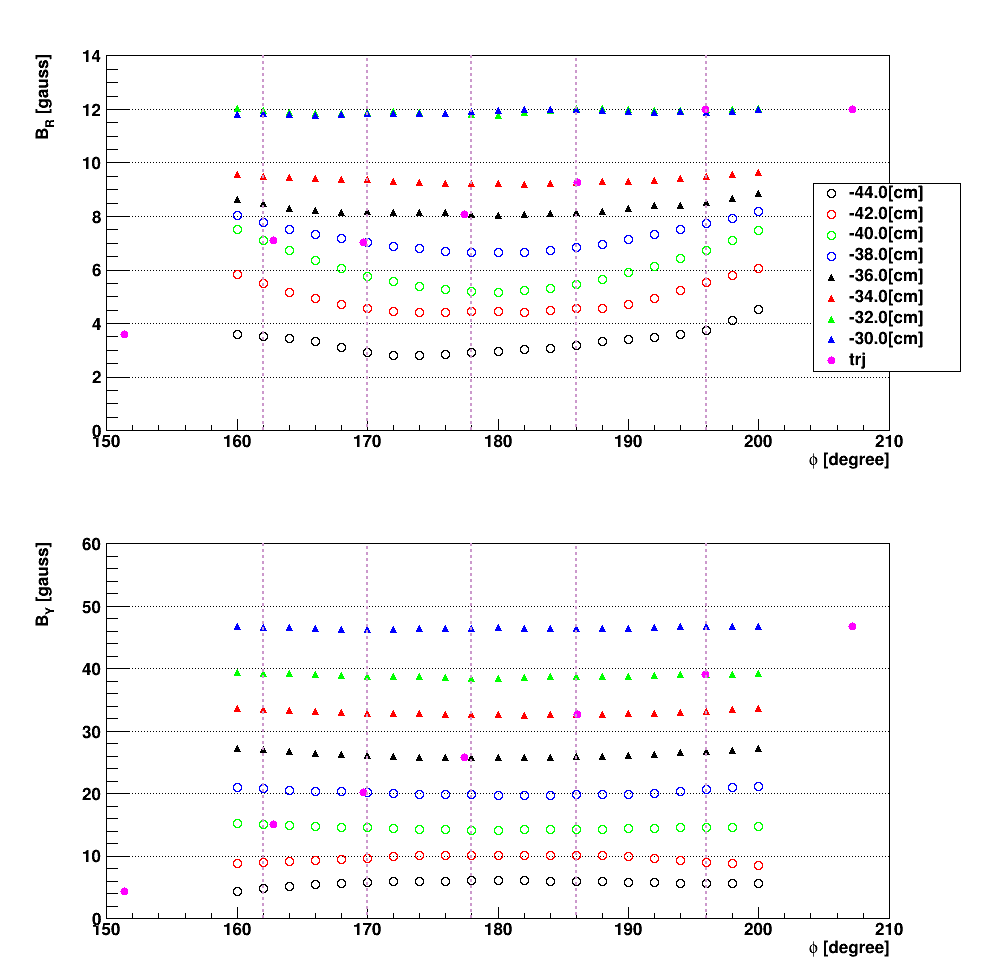}
\caption{
Azimuthal dependence of the magnetic field in the non-axisymmetric injection sector. Different curves represent different positions along the solenoid axis \(Y\), and the magenta points indicate the approximate locations sampled by the injection trajectory. A pronounced azimuthal dependence is observed owing to the non-axisymmetric geometry of the iron yoke.}
   \label{fig:SectorField}
\end{figure}

\begin{figure}[!htb]
   \centering  
   \includegraphics*[width=.8\columnwidth]{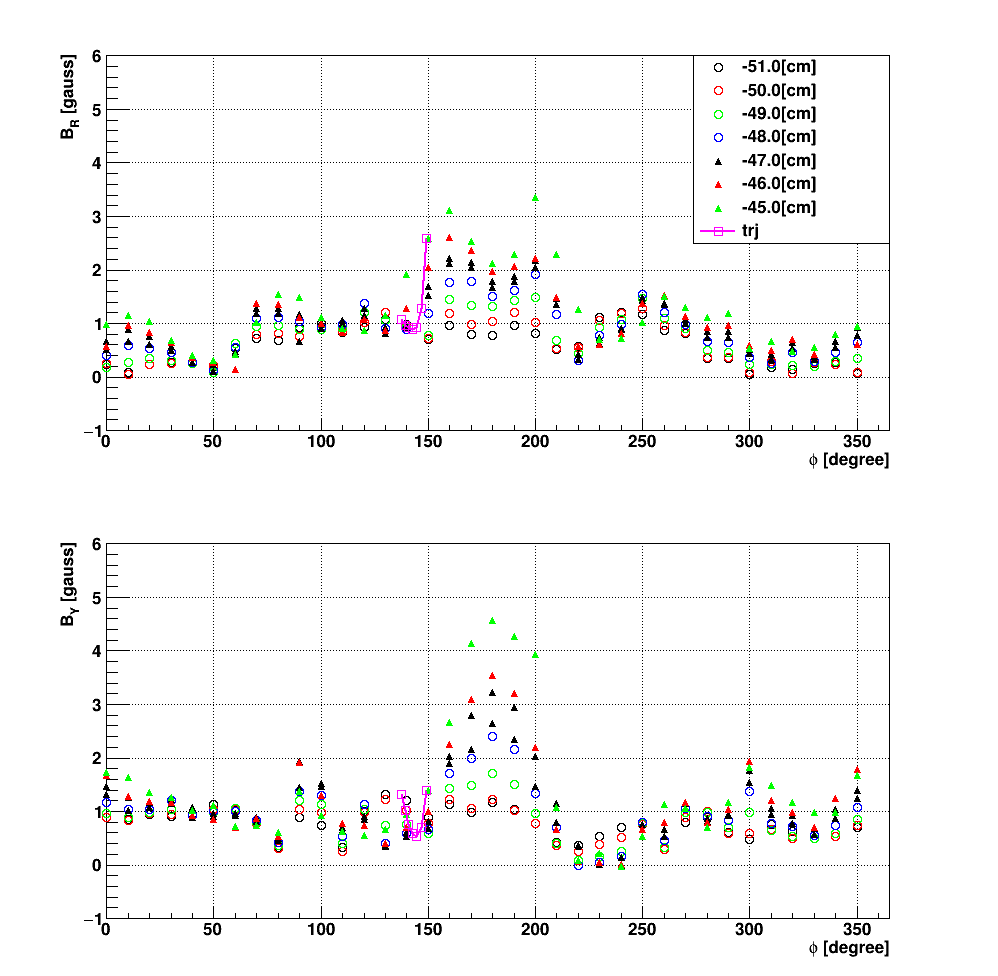}
   \caption{
   Azimuthal dependence of the magnetic field in the region outside the iron yoke. Different curves represent different positions along the solenoid axis \(Y\). The field exhibits an azimuthal dependence, although its magnitude and variation are smaller than those in the injection sector.}
   \label{fig:OuterField}
\end{figure}

Figures~\ref{fig:SectorField} and ~\ref{fig:OuterField} 
show the azimuthal dependence of the magnetic field in the sector-shaped injection region and in the region outside the iron yoke, respectively.
 The magenta points indicate the approximate locations sampled by the injection trajectory. In the injection sector, the magnetic field exhibits a pronounced azimuthal dependence owing to the non-axisymmetric geometry of the iron yoke. Consequently, azimuthal averaging over the entire sector would smear out the field structure relevant to the particle motion. For the reconstruction, it is therefore more appropriate to retain the local field distribution in the limited azimuthal region traversed by the injected particles.

Based on this observation, the injection sector was divided into narrow azimuthal slices, and the equivalent-current reconstruction was performed independently for each slice. In this way, the local azimuthal dependence of the magnetic field relevant to the injection trajectory is retained instead of being removed by azimuthal averaging.

The target field components used in the tSVD reconstruction are selected according to the symmetry of each region. In the approximately axisymmetric region inside the iron yoke, the radial and axial components, \(B_R\) and \(B_Y\), are used as the reconstruction targets. The reconstructed radial component is evaluated from the Cartesian field components as
\begin{equation}
 B_R^{\mathrm{calc}} =\frac{xB_x^{\mathrm{calc}}+zB_z^{\mathrm{calc}}}{R}. 
 \end{equation}

In contrast, the axial symmetry is significantly broken in the injection sector. The local field components are therefore retained directly, and the equivalent currents are determined by a weighted two-component fit,

\begin{equation}
    \chi^2= \sum_i\left[ w_R^2\left(B_{R,i}^{\mathrm{target}}-B_{R,i}^{\mathrm{rec}}\right)^2+ w_Y^2\left(B_{y,i}^{\mathrm{target}}-B_{y,i}^{\mathrm{rec}}\right)^2 \right].
\end{equation}  

This corresponds to solving the weighted linear system
\begin{equation}
 \begin{pmatrix} w_R\mathbf B_R\\ w_Y\mathbf B_y \end{pmatrix} = \begin{pmatrix} w_R\mathbf M_R\\ w_Y\mathbf M_y \end{pmatrix} \mathbf I 
\end{equation}
by tSVD.
Thus, both the spatial sampling and the target field components are adapted to the local magnetic-field symmetry.

Once the equivalent-current amplitudes \(\mathbf{I}\) are determined, both the magnetic field \(\mathbf{B}\) and the vector potential \(A_\phi\) are calculated from the same current elements. This provides the vector potential corresponding to the reconstructed magnetic-field distribution.

\section{VALIDATION OF THE MAGNETIC-FIELD RECONSTRUCTION}
\begin{figure}[!htb]
   \centering
   \includegraphics*[width=0.9\columnwidth]{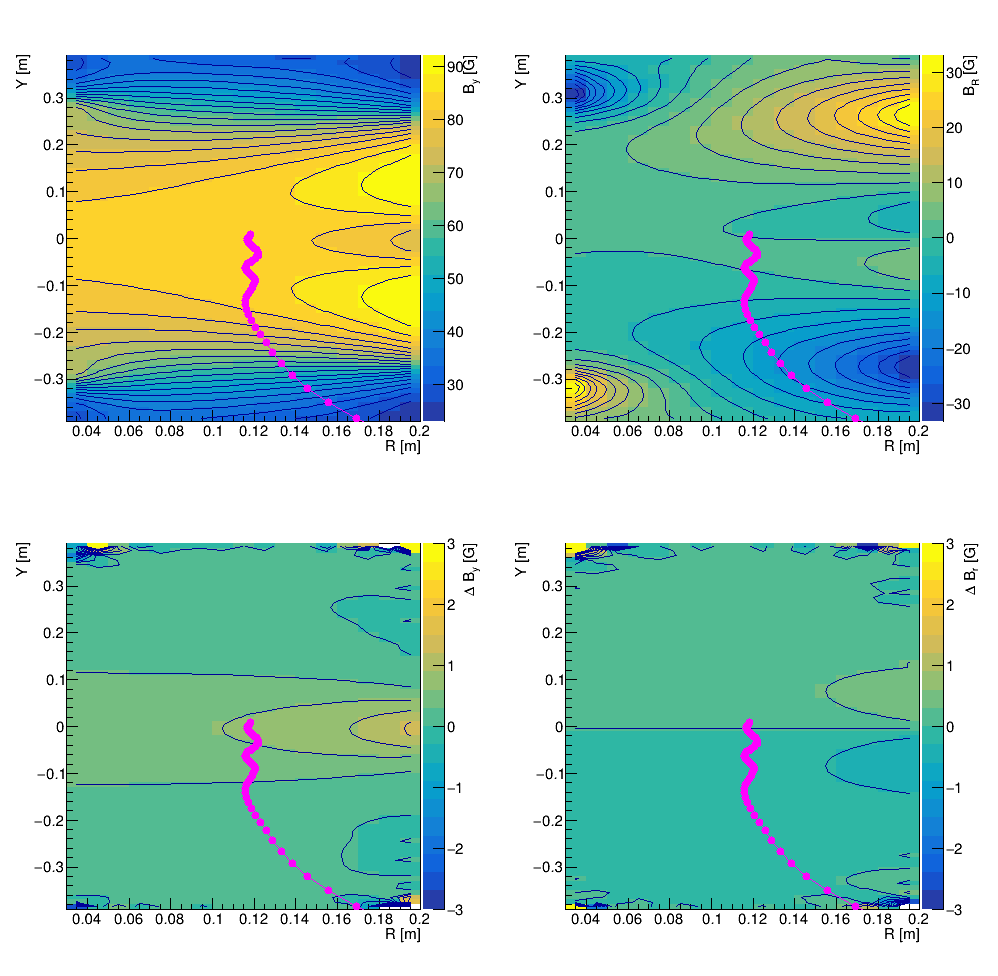}
   \caption{
   Validation of the magnetic-field reconstruction in the non-axisymmetric injection sector. The upper panels show the target magnetic-field distributions, and the lower panels show the differences between the target and reconstructed fields, \(\Delta B_k=B_k^{\mathrm{target}}-B_k^{\mathrm{rec}}\). The magenta curve and points indicate the representative injection trajectory and the sampled positions along it. The white region is excluded from the comparison because it corresponds to the equivalent-current model region.}
   \label{fig:ReconInnerField}
\end{figure}

\begin{figure}[!htb]
   \centering
   \includegraphics*[width=0.9\columnwidth]{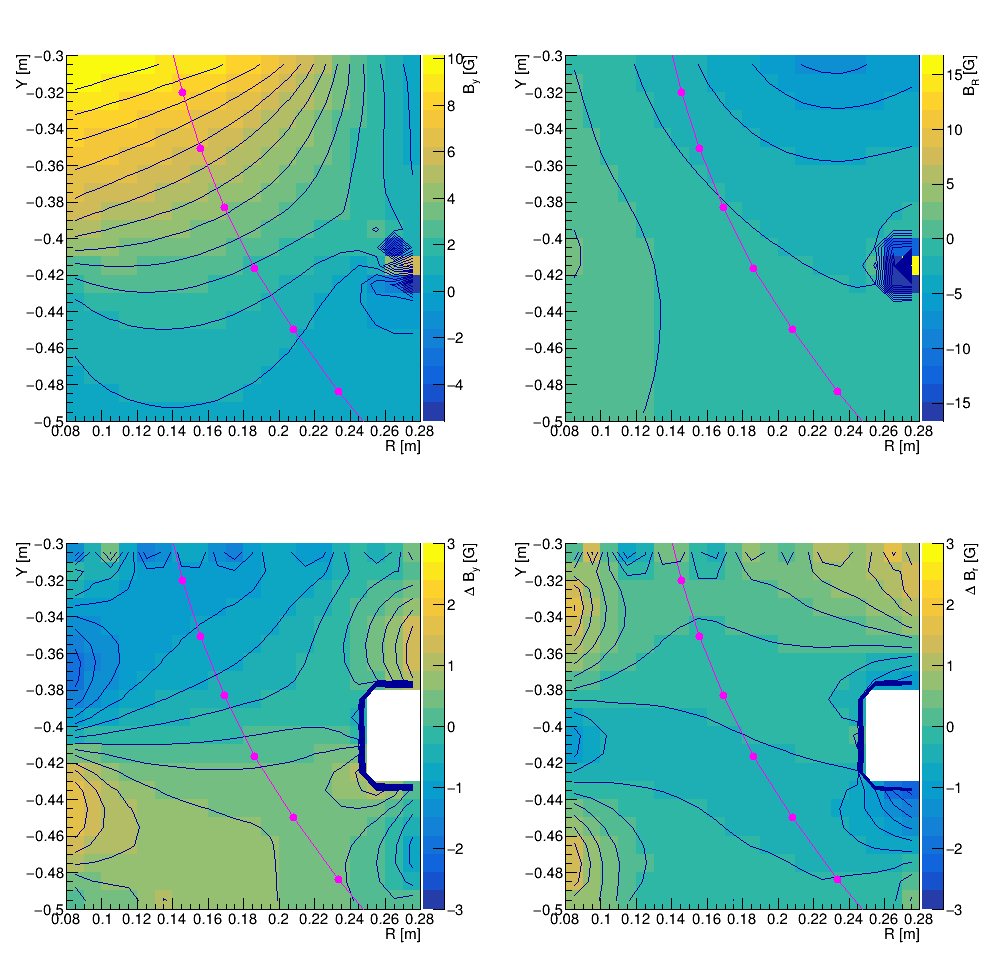}
   \caption{Validation of the magnetic-field reconstruction in the approximately axisymmetric region inside the iron yoke. The upper panels show the azimuthally averaged target fields \(B_Y\) (left) and \(B_R\) (right), and the lower panels show the corresponding differences between the target and reconstructed fields, \(\Delta B_Y\) and \(\Delta B_R\). The magenta curve indicates the representative injection trajectory.
   }
   \label{fig:ReconSectorField}
\end{figure}

The accuracy of the equivalent-current reconstruction was evaluated by direct comparison with the target three-dimensional magnetic-field map. The field difference is defined as

\begin{equation}
 \Delta B_k = B_k^{\mathrm{target}}-B_k^{\mathrm{rec}}, 
\end{equation}
where \(k\) denotes the field component used in each reconstruction region. Figures~\ref{fig:ReconInnerField} and ~\ref{fig:ReconSectorField} show representative comparisons for the non-axisymmetric injection sector and the approximately axisymmetric region inside the iron yoke, respectively.

Figure~\ref{fig:ReconInnerField} shows the reconstruction in the injection sector, where the local field components are fitted without azimuthal averaging. Figure~\ref{fig:ReconSectorField} shows the corresponding result inside the iron yoke, where the azimuthally averaged \(B_R\) and \(B_Y\) distributions are used as the reconstruction targets. In both figures, the upper panels show the target magnetic-field distributions and the lower panels show the differences between the target and reconstructed fields. The magenta curve indicates the representative injection trajectory.

The reconstructed fields reproduce the overall spatial distributions of the target fields in both the non-axisymmetric injection sector and the approximately axisymmetric region. The residual distributions are substantially smaller than the corresponding field distributions over most of the reconstruction domain. Localized deviations are observed near the boundary of the equivalent-current model, particularly in regions where the magnetic field varies rapidly.

Along the representative injection trajectory, the residuals remain relatively small and vary smoothly in both regions. These results indicate that the equivalent-current reconstruction retains the magnetic-field structure relevant to the particle motion, including the non-axisymmetric field variation in the injection sector. The reconstructed current distribution can therefore be used to evaluate the corresponding vector potential \(A_\phi\) along the injection trajectory.

The reconstruction was also examined in the region outside the iron yoke, where a weaker azimuthal dependence remains as shown in Fig.~\ref{fig:OuterField}. A comparable level of agreement between the target and reconstructed fields was obtained; therefore, only the results for the injection sector and the inner approximately axisymmetric region are shown here.

\section{EVALUATION OF THE VECTOR POTENTIAL AND CANONICAL ANGULAR MOMENTUM}

The axisymmetric magnetic-field representation described in the previous section provides the azimuthal vector potential \(A_\phi\), from which the canonical angular momentum can be constructed. 
For an axisymmetric magnetic field, the vector potential is expressed as

\begin{equation}
    \vec{A}=(0,A_\phi, 0)
\end{equation}
where the radial and axial components vanish under the adopted gauge. The canonical momentum is defined by

\begin{equation}
    \vec{P}=\vec{p}+q\vec{A}
\end{equation}

where \(\mathbf{p}\) denotes the mechanical momentum and \(q\) is the particle charge.

\begin{figure}[!htb]
   \centering
   \includegraphics*[width=0.9\columnwidth]{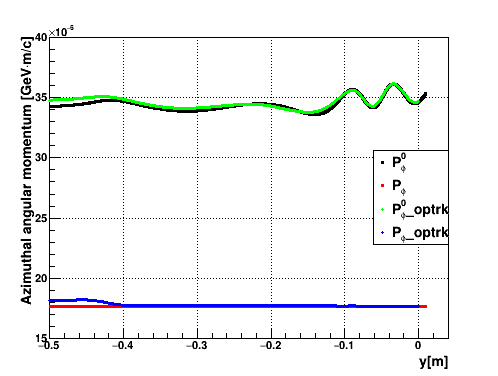}
   \caption{
   Mechanical \(P_\phi^0\) and canonical \(P_\phi\) angular momenta along the OPERA trajectory (green and blue) and the axisymmetric tSVD trajectory (black and red), respectively, calculated with identical initial conditions. The canonical angular momentum is conserved in the axisymmetric model and remains approximately constant along the OPERA trajectory.}
   \label{fig:tmp}
\end{figure}

With the azimuthal angle defined around the \(Y\) axis, the mechanical and canonical angular momenta are written as

\begin{equation}
 P_\phi^{0}=xp_z-zp_x , 
\end{equation}
and
\begin{equation}
 P_\phi=P_\phi^{0}+qRA_\phi , 
\end{equation}
respectively.
Figure~\ref{fig:tmp} compares the mechanical and canonical angular momenta evaluated along the OPERA trajectory and the trajectory calculated using the axisymmetric tSVD field model with identical initial conditions. 
The mechanical angular momenta agree closely, while the canonical angular momentum \(P_\phi\) is conserved in the axisymmetric model and remains approximately constant along the OPERA trajectory.

These results support the applicability of the reconstructed vector potential for evaluating canonical angular momentum in the approximately axisymmetric region. 

In an axisymmetric magnetic field, the conservation of canonical angular momentum constrains injection trajectory design, through which the spatial distribution of the vector potential is reflected in the resulting particle motion.



\section{CONCLUSION}
A method for reconstructing the vector potential from a three-dimensional magnetic-field map was developed for the canonical description of three-dimensional spiral injection. Equivalent-current models were constructed according to the local magnetic-field symmetry, and the reconstructed fields were shown to reproduce the reference field distributions in both the approximately axisymmetric region and the non-axisymmetric injection sector.

The reconstructed vector potential was then used to evaluate the canonical angular momentum along the representative injection trajectory. In the approximately axisymmetric region, the inclusion of the vector-potential contribution substantially reduced the variation observed in the mechanical angular momentum, resulting in an approximately constant canonical angular momentum. This demonstrates the effectiveness of the canonical description for characterizing particle motion directly in a spatially varying magnetic field.

Further work will extend the evaluation to the non-axisymmetric injection region and examine the consistency of the vector-potential reconstruction across independently reconstructed field regions. This will provide a basis for assessing the applicability of canonical angular momentum as a practical design parameter in arbitrary magnetic-field configurations.

\end{document}